# Thermal conductivity of tetrahydrofuran hydrate

A. I. Krivchikov[a], V.G. Manzhelii[a], O.A. Korolyuk[a], B.Ya. Gorodilov[a], O.O. Romantsova[a]

[a] *B.Verkin Institute for Low Temperature Physics and Engineering of NAN Ukraine, 61103, 47 Lenin Ave., Kharkov, Ukrain. Fax: 38 057 340 3370; Tel: 38 057 3408 582; E-mail: krivchikov@ilt.kharkov.ua*

*This submission was created using the RSC Article Template (DO NOT DELETE THIS TEXT)*
*(LINE INCLUDED FOR SPACING ONLY - DO NOT DELETE THIS TEXT)*

The thermal conductivity of tetrahydrofuran hydrate has been measured in the temperature region 2 – 220 K by the steady-state potentiometric method. The temperature dependence of the thermal conductivity exhibits behavior typical of amorphous substances. It is shown that above 100 K the mean free path of the phonons is considerably smaller than the lattice parameter and is no longer dependent on temperature.

## I. Introduction

Clathrate hydrates (the chemical formula is $G \cdot zH_2O$, where G is the Van der Waals-type substance and z is hydrate number, $z = 5\frac{2}{3} \div 34$) consist of water molecules, forming rather large polyhedral cavities, and guest particles residing in these cavities [1]. The guest particles can be inert gas atoms or small molecules. Clathrate hydrate can be viewed as a model system for investigating the origin of the glass-like behavior of the thermal conductivity in solids having a crystal structure [2,3,4]. Most clathrate hydrates crystallize into one of the two cubic structures with the unit cells larger than that of ordinary ice Ih. The size of the guest molecules dictates the type of the crystal structure of clathrate hydrates. Most of the physical properties of clathrate hydrates are similar to those of ice Ih. The exception is the unusually low thermal conductivity whose temperature dependence is unnatural in crystal substances. In clathrate hydrates with different guest molecules (methane, xenon, ethylene oxide, propane, tetrahydrofuran $C_4H_8O$ (THF), 1,3 – dioxolane, cyclobutane) the thermal conductivity, which is only weakly dependent on the nature of guest molecules, behaves like in glass substances [2,3,5-8]. Recently, such behavior of thermal conductivity has been detected in doped semiconducting clathrates [9-11]. The typical feature of clathrate compounds is the presence of low-lying optical modes in vibrational energy spectrum, which interact intensively with the acoustic modes. The predicted [12,13] strong optical-acoustic mode coupling induced by the Van der Waals interaction between the guest molecule and the water molecules was supported in experiments on inelastic neutron scattering [14-16] and by inelastic X-ray scattering [17] in methane and xenon hydrates

In this study the thermal conductivity of tetrahydrofuran $C_4H_8O$ (THF) hydrate (THF·17H$_2$O) was measured at low temperatures. The THF hydrate crystallizes into the type II – cubic structure (Fd3m, a=17.1 Å). Its stoichiometric structure is $16X \cdot 8Y \cdot 136H_2O$ [1]. For each unit cell containing 136 water molecules there are eight large hexakaidecahedra and 16 small dodecahedra. Each large hexakaidecahedron contains one THF molecule. The small dodecahedra are all empty. Unlike other hydrates, the THF hydrate is obtained by crystallization of a liquid water – THF solution with a molar ratio of 17/1. The THF hydrate can have two types of crystal disorder. The proton disorder is due to the arrangement of hydrogen atoms in the crystal lattice formed by water molecules. The proton disorder appears when the spatial arrangement of protons becomes frozen at the glass transition temperature, T = 82 K [18,19]. The orientational disorder is produced by the rotational motion of guest THF molecules and persists down to 20 K [20].

The thermal conductivity of the THF hydrate was investigated earlier by several research groups [2,3,8,21-23]. The first measurement [2] of thermal conductivity in an interval of 100 – 260 K showed its unusual dependence on temperature. At lowering temperature the thermal conductivity was observed to decrease in contrast to its growth in a dielectric crystal. The glass-like behavior of the thermal conductivity of THF at lower (15 – 100 K) temperatures was reported in [3].

This study was motivated by the following. Firstly, there were no experimental data about the thermal conductivity of the THF hydrate at temperatures below 15 K. But it is precisely in this region that the temperature dependence of thermal conductivity is expected to change. Secondly, a 3-5 times variance in literature data prohibited unambiguous conclusions about the behavior of thermal conductivity in a wide range of temperatures. Thirdly, the investigations on THF hydrate promise more complete and reliable experimental information about thermal conductivity: in contrast to other clathrate hydrates, e.g., Xe or CH$_4$ hydrates [24], it is a relatively simple matter to prepare a high-quality homogeneous solid THF hydrate sample for measuring its thermal conductivity by the standard steady-state potentiometric method.

## II. Experimental Results

A THF·16.9H$_2$O sample was prepared from 99.9% pure tetrahydrofuran (ACROS Organics) and double-distilled water. The THF components and water were mixed in the ratio of 1:16.9. The stoichiometric coefficient was chosen to be slightly below 17 to avoid the formation of crystal ice inclusions in the sample. The prepared homogeneous liquid mixture was put in the container of the measuring cell. The basic diagram of the measuring cell is shown in Fig.1.

The sample container (1) is a stainless steel 40 mm long tube 22 mm in diameter with the wall thickness 0.3 mm. The container bottom was fixed in the cooled zone of the cryostat (4) connected to a helium bath (7). Two copper wires (8) 1 mm in diameter were passed through the container perpendicular to its axis, which permitted measurement of the average temperature along the isothermal plane running across the sample. The wires were 12.3 mm apart along the container axis. At the outer surface of the container copper sockets were soldered to the wires (8) to cartridge two temperature sensors. The upper sensor is a Cernox – SD resistance thermometer (2) (Lake Shore Cryotronics, Inc.) measuring the temperature difference; the lower sensor is a TSU-2 resistance thermometer (3) (VNIIFTRI) used to stabilize and control the temperature. Thermometer (2) was used to measure the temperature along the sample with the heat flow on and off. The container with the sample was vacuum-tight covered with a copper In-sealed cap (5).

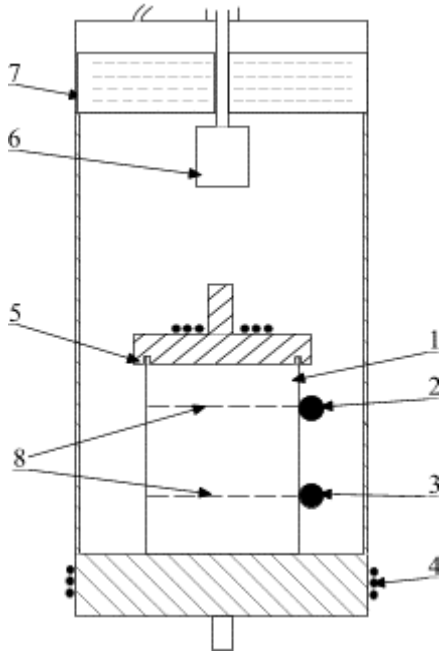

Fig.1. The diagram of the cell for the measuring thermal conductivity: 1 – container for sample, 2 – Cernox resistance thermometer, 3 – TSU-2 resistance thermometer, 4 – bottom block (cooler) with a heater, 5 – cap with a heater, 6 – block for electric leads-in, 7 – helium bath, 8 – copper wire.

A heater was mounted on the cap (5) to generate a heat flow over the sample.

In the container the clathrate hydrate was grown during 30 min. by solidifying the liquid mixture at T = 278.15 K with cooling the bottom zone of the container.

The thermal conductivity of the THF hydrate was measured in the interval 2 – 220 K by the steady-state potentiometric method.

The temperature dependence of the thermal conductivity coefficient $k(T)$ of THF hydrate obtained in this study is shown in Fig.2 along with literature data. Our results agree well with literature data, except for [3].

As see in Fig.2, the temperature dependence of the thermal conductivity coefficient of THF has the shape typical of amorphous solids: it is close to quadratic at liquid helium temperatures and has a "plateau" (where the thermal conductivity is practically temperature independent) as the temperature increases. It should be noted, however, that this two features are observed at higher temperatures as compared to amorphous solids [26]. In the region of the proton glass transition T = 85 K, the thermal conductivity curve has a smeared maximum. Its position coincides with the temperature of the maximum in the thermal conductivity curve of THF hydrate containing KOH impurity [8]. The KOH impurity speeds up the motion of the hydrogen atoms between the equivalent positions.

## III. Discussion

According to the simple gas-kinetic model of thermal conductivity,

$$k(T) = 1/3\, C_{ph}\, v\, l, \qquad (1)$$

where $C_{ph}$ is the phonon heat capacity, $v$ is the mean velocity of phonon propagation, $l$ is the mean free path of the phonons.

The temperature dependence of the phonon mean free path calculated by Eq. (1) from our thermal conductivity data using the heat capacity calculated in the Debye model for the Debye temperature $\Theta_D$ = 198 K and $v$ = 1871 m/s [3] is shown in Fig.3. Two temperature regions are evident, in which the mechanisms of heat transfer are fundamentally different. At low temperatures the motion of the phonons is ballistic, i.e. their mean free path exceed considerably the distance a = 2.78 Å between the oxygen atoms forming the clathrate framework. As the temperature rises, the phonon mean free path $l$ decreases and becomes temperature-independent above 100 K reaching the value 4 ÷ 5 Å.

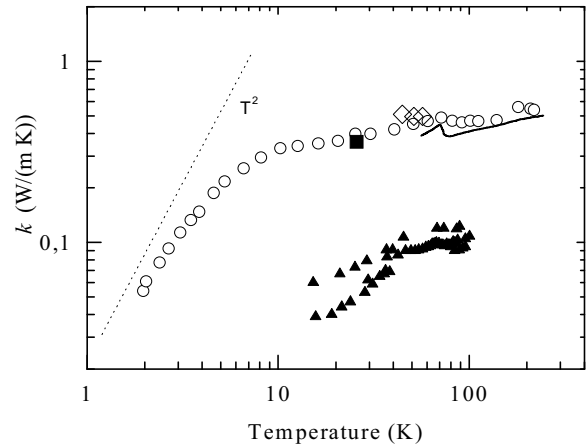

Fig.2. Thermal conductivity of THF clathrate hydrate and literature data: ○ - our data, ▲ - [3], —— - [8], ◊ - [23], ■ - [25].

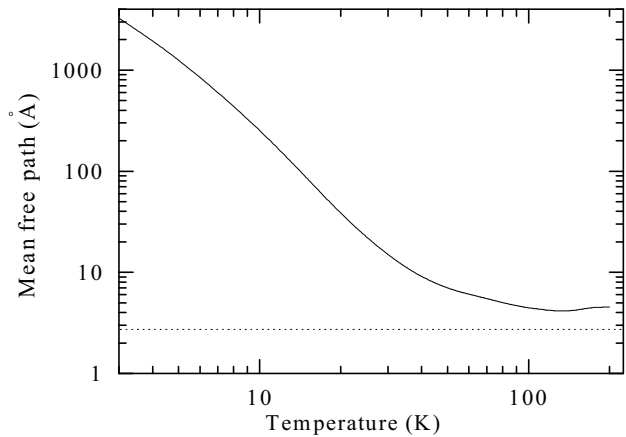

Fig.3. The dependence of the phonon mean free path on temperature (solid curve) calculated from the thermal conductivity data. The mean distance between the oxygen atoms in the THF hydrate lattice is shown by the dotted curve.

This behavior of $l$ is characteristic of the diffusive motion of phonons (the Ioffe-Riegel regime). In the strict sense, the concept of phonons is unusable with such low values of $l$, and the thermal conductivity is effected by a direct energy exchange between the neighboring molecules [27,28].

The change from the ballistic motion to the diffusive one occurs in a wide (~ 30 – 100 K) interval of temperatures, where both the mechanisms of heat transfer are in competition. The contribution of the phonons to the one or the other heat transfer regime is dependent on their wavelengths [29].

The glass-like behavior of the thermal conductivity of THF hydrate at low and high temperatures may be attributed first of all to the interaction between the acoustic phonons and the local vibrations of the guest THF molecules. In the clathrate structure the vibrations of the guest and host molecules are strongly coupled. The guest molecules residing inside the clathrate cavity are coupled to it by weak Van der Waals forces and their vibrational motions is anharmonic. The H-bonded framework of the water molecules shifts from the positions of equilibrium and "pushes" the guest atoms [15]. On the other hand, vibrations of the guest molecules has influence on the lattice vibrations of the hydrate, i.e. the hydrate vibrations are modulated by the vibrations of the guest molecules [12,17]. In this case the ballistic motions is feasible for the phonons whose energy is smaller than that of the local vibrations. Three kinds of local vibrations with 1.8, 2.8 and 10 meV occur in the THF hydrate [30]. The acoustic phonons having energies equal to or high than the energy of the local vibrations are scattered strongly by this vibrations. Because of the strong scattering of the phonons by the guest molecules, the motion of the acoustic phonons over the crystal becomes diffusive.

## VI. Conclusions

Thus, our experimental results on the thermal conductivity of the THF hydrate in the temperature region 2 – 220 K show that the behavior of the temperature conductivity is typical of amorphous substances. Above 100 K the phonon mean free path is considerably smaller than the lattice parameter of the hydrate and is no longer dependent on temperature. The motion of acoustic phonons becomes diffusive.

## Acknowledgents

This work has been partially funded as a project by Forschungszentrum Jülich. We thank Prof. Werner Press, Dr Harald Conrad and Dr H. Schober for formulating the problem, a stimulated discussion and their continued interest in the work.

## References

1 Julian Baumert, *Structure, Lattice Dynamics, and Guest Vibrations of Methane and Xenon Hydrate,* Dissertation, zur Erlangung des Doktorgrades der Mathematisch-Naturwissenschaftlichen Fakult at der Christian-Albrechts-Universit at zu Kiel 2003
2 R.G. Ross, P. Andersson, G. Backström, *Nature* 1981, **290**, 322.
3 J. S. Tse, Mary Anne White, *J.Phys.Chem.,* 1988, **92**, 5006.
4 J.S. Tse, M.L. Klein, and I.R. McDonald, *J. Chem. Phys.* 1983, **78**, 2096.
5 J. G. Cook, D. G. Leaist, *Geophys. Res. Lett.,* 1983, **10**, 397.
6 P. Andersson, R. G. Ross, *J. Phys. C,* 1983, **16**, 1423.
7 N. Ahmad, W. A. Phillips, *Solid State Comm.*, 1987, **63**, 167.
8 O. Andersson and H. Suga, *J. Phys. Chem. Solids*, 1996, 57, 1, 125.
9 J.S. Tse, K. Uehara, R. Rousseau, A. Ker, C.I. Ratcliffe, M.A. White, G. McKay, *Phys. Rev. Lett.*, 2000, **85**, 114.
10 B. C. Sales, B. C. Chakoumakos, R. Jin, J. R. Thompson, and D. Mandrus, *Phys. Rev. B,* 2001, **63**, 245113.
11 A. Bentien, M. Christensen, J. D. Bryan, A. Sanchez, S. Paschen, F. Steglich, G. D. Stucky, and B. B. Iversen, *Phys. Rev. B,* 2004, **69**, 045107.
12 R. Inoue, H. Tanaka, K. Nakanishi, *J. Chem. Phys.*, 1996, **104** 9569.
13 J.S. Tse, V.P. Shpakov, V.V. Murash ov, V.R. Belosludov, *J. Chem. Phys.,* 1997, **107**, 9271.
14 J.S. Tse, V.P Shpakov, V.R. Belosludov, F. Trouw, Y.P., Handa, and W. Press, *Europhys. Lett.*, 2001, **54**, 354.
15 C. Gutt, J. Baumert, W. Press, J.S. Tse, S. Janssen, *J. Chem. Phys.* 2002, **116**, 3795.
16 H. Schober, H. Itoh, A. Klapproth, V. Chihaia, and W.F. Kuhs, *E. J. Phys. E*, 2003, **12**, 41.
17 J. Baumert, C. Gutt, V.P. Shpakov, J.S. Tse, M. Krisch, M. Muller, H. Requardt, D.D. Klug, S. Janssen, and W. Press, *Phys. Rev. B*, 2003, **68**, 174301.
18 O. Yamamuro, M. Oguni, T. Matsuo and H. Suga, *Solid St. Comm,* 1987, **62**, 289.
19 O. Yamamuro, M. Oguni, T. Matsuo and H. Suga, *J. Phys. Chem. Solids,* 1988, **49**, 425.
20 T. M. Kirschgen, M. D. Zeidler, B. Geil and F. Fujara, *Phys. Chem. Chem. Phys.*, 2003, **5**, 5243.
21 R.G. Ross, P. Andersson, *Can. J. Chem*. 1982, **60**, 881.
22 J. G. Cook and M. J. Laubitz *in Thermal Conductivity*, ed. J. G. Hust, Plenum Press, New York, 1983, Vol. 17, p. 745.
23 T. Ashworth, L. R. Johnson and L. –P. Lai, *High Temp. High Pressure* 1985, **17**, 413.
24 W. F. Kuhs, A. Kalpproth, F. Gotthardt, K. Techmer, T.Heinrichs, *Geophys. Res. Lett.,* 2000, **27**, 2929.
25 S.Kulikov et al., *Radiation Effects in Cold Moderator Materials: Experimental Study of Accumluation and Release of Chemical Energy*, Proceedings of the 16th Meeting of the International Collaboration on Advanced Neutron Sources, Neuss, Germany,May 2003.
26 D. G. Cahill, R.O. Pohl, *Ann. Rev. Chem*., 1988, **39**,93.
27 D.G. Cahill, S. K. Watson, and R.O. Pohl, *Phys.Rev.B.,* 1992, **46**, 6131.
28 Ph. B. Allen and J. L. Feldman, *Phys. Rev. B*, 1993, **48**, 12581.
29 V.A. Konstantinov, *Low Temp. Phys*., 2003, **29**, 422.
30 H. Conrad, W. F. Kuhs, K. Nünighoff, C. Pohl, M. Prager, *Physica B* to be published in 2004